\documentclass[aps,prd,preprint,groupedaddress]{revtex4-1}

\usepackage{amsmath,amssymb,amsfonts}
\usepackage{graphicx}
\usepackage{array}
\newcolumntype{L}{>{\centering\arraybackslash}p{3cm}}

\begin{document}

\title{Kaon oscillations and baryon asymmetry of the universe}


\author{Wanpeng Tan}
\email[]{wtan@nd.edu}
\affiliation{Department of Physics, Institute for Structure and Nuclear Astrophysics (ISNAP), and Joint Institute for Nuclear Astrophysics - Center for the Evolution of Elements (JINA-CEE), University of Notre Dame, Notre Dame, Indiana 46556, USA}

\date{\today}

\begin{abstract}
Baryon asymmetry of the universe (BAU) can likely be explained with $K^0-K^{0'}$ oscillations of a newly developed mirror-matter model and new understanding of quantum chromodynamics (QCD) phase transitions. A consistent picture for the origin of both BAU and dark matter is presented with the aid of $n-n'$ oscillations of the new model. The global symmetry breaking transitions in QCD are proposed to be staged depending on condensation temperatures of strange, charm, bottom, and top quarks in the early universe. The long-standing BAU puzzle could then be understood with $K^0-K^{0'}$ oscillations that occur at the stage of strange quark condensation and baryon number violation via a non-perturbative sphaleron-like (coined ``quarkiton'') process. Similar processes at charm, bottom, and top quark condensation stages are also discussed including an interesting idea for top quark condensation to break both the QCD global $U_t(1)_A$ symmetry and the electroweak gauge symmetry at the same time. Meanwhile, the $U(1)_A$ or strong $CP$ problem of particle physics is addressed with a possible explanation under the same framework. 
\end{abstract}

\pacs{}

\maketitle

\section{Introduction\label{intro}}

The matter-antimatter imbalance or baryon asymmetry of the universe (BAU) has been a long standing puzzle in the study of cosmology. Such an asymmetry can be quantified in various ways. The cosmic microwave background (CMB) data by Planck set a very precise observed  baryon density of the universe at $\Omega_bh^2 = 0.02242\pm0.00014$ \cite{planckcollaboration2018}. This corresponds to today's baryon-to-photon number density ratio of $n_B/n_\gamma = 6.1\times10^{-10}$. For an adiabatically expanding universe, it would be better to use the baryon-number-to-entropy density ratio of $n_B/s = 8.7\times10^{-11}$ to quantify the BAU, which may have to be modified under the new understanding of the neutrino history in the early universe (see Sec. \ref{origin}).

From known physics, it is difficult to explain the observed BAU. For example, for an initially baryon-symmetric universe, the surviving 
relic baryon density from the annihilation process is about nine orders of magnitude lower than the observed one \cite{kolb1990}. Therefore, an asymmetry is needed in the early universe and the BAU has to exist before the temperature of the universe drops below $T=38$ MeV \cite{kolb1990} to avoid the annihilation catastrophe between baryons and anti-baryons.

Sakharov proposed three criteria to generate the initial BAU: (i) baryon number (B-) violation (ii) $C$ and $CP$ violation (iii) departure from thermal equilibrium \cite{sakharov1967}. The Standard Model (SM) is known to violate both $C$ and $CP$ and it does not conserve baryon number only although it does $B-L$ (difference of baryon and lepton numbers). Coupled with possible non-equilibrium in the thermal history of the early universe, it seems to be easy to solve the BAU problem. Unfortunately, the violations in SM without new physics are too small to explain the observed fairly large BAU. The only known B-violation processes in SM are non-perturbative, for example, via the so-called sphaleron \cite{klinkhamer1984} which involves nine quarks and three leptons from each of the three generations. It was also found out that the sphaleron process can be much faster around or above the temperature of the electroweak symmetry breaking or phase transition $T_{\text{EW}} \sim 100$ GeV \cite{kuzmin1985}. This essentially washes out any BAU generated early or around $T_{\text{EW}}$ since the electroweak transition is most likely just a smooth cross-over instead of ``desired'' strong first order \cite{kajantie1996}. It makes the appealing electroweak baryogenesis models \cite{kuzmin1985,farrar1993} ineffective and new physics often involving the Higgs have to be added in the models \cite{trodden1999,cline2006,cline2018b}. Recently lower energy baryogenesis typically using particle oscillations stimulated some interesting ideas \cite{elor2019,bringmann2019}. Other types of models such as leptogenesis \cite{buchmuller2005} are typically less testable or have other difficulties.

Here we present a simple picture for baryogenesis at energies around quantum chromodynamics (QCD) phase transition with $K^0-K^{0'}$ oscillations based on a newly developed mirror matter model \cite{tan2019}. $K^0-K^{0'}$ oscillations and the new mirror matter model will be first introduced to demonstrate how to generate the ``potential'' amount of BAU as observed. Then the QCD phase transition will be reviewed and the sphaleron-like non-perturbative processes are proposed to provide B-violation and realize the ``potential'' BAU created by $K^0-K^{0'}$ oscillations. In the end, the observed BAU is generated right before the $n-n'$ oscillations that determine the final mirror(dark)-to-normal matter ratio of the universe \cite{tan2019}. Meanwhile, the long-standing $U(1)_A$ and strong $CP$ problems in particle physics are also naturally resolved under the same framework.

\section{$K^0-K^{0'}$ oscillations and the new model\label{KK}}

To understand the observed BAU, we need to apply the newly developed particle-mirror particle oscillation model \cite{tan2019}. It is based on the mirror matter theory \cite{kobzarev1966,blinnikov1983,kolb1985,khlopov1991,hodges1993,foot2004,berezhiani2004,okun2007}, that is, two sectors of particles have identical interactions within their own sector but share the same gravitational force. Such a mirror matter theory has appealing theoretical features. For example, it can be embedded in the $E_8\otimes E_{8'}$ superstring theory \cite{green1984,gross1985,kolb1985} and it can also be a natural extension of recently developed twin Higgs models \cite{chacko2006,barbieri2005} that protect the Higgs mass from quadratic divergences and hence solve the hierarchy or fine-tuning problem. The mirror symmetry or twin Higgs mechanism is particularly intriguing as the Large Hadron Collider has found no evidence of supersymmetry so far and we may not need supersymmetry, at least not below energies of 10 TeV. Such a mirror matter theory can explain various observations in the universe including the neutron lifetime puzzle and dark-to-baryon matter ratio \cite{tan2019}, evolution and nucleosynthesis in stars \cite{tan2019a}, ultrahigh energy cosmic rays \cite{tan2019b}, dark energy \cite{tan2019e}, and a requirement of strongly self-interacting dark matter to address numerous discrepancies on the galactic scale \cite{spergel2000}.

In this new mirror matter model \cite{tan2019}, no cross-sector interaction is introduced, unlike other particle oscillation type models. The critical assumption of this model is that the mirror symmetry is spontaneously broken by the uneven Higgs vacuum in the two sectors, i.e., $<\phi> \neq <\phi'>$, although very slightly (on a relative breaking scale of $\sim 10^{-15} \text{--} 10^{-14}$) \cite{tan2019}. When fermion particles obtain their mass from the Yukawa coupling, it automatically leads to the mirror mixing for neutral particles, i.e., the basis of mass eigenstates is not the same as that of mirror eigenstates, similar to the case of ordinary neutrino oscillations due to the family or generation mixing. The Higgs mechanism makes the relative mass splitting scale of $\sim 10^{-15} \text{--} 10^{-14}$ universal for all the particles that acquired mass from the Higgs vacuum. Further details of the model can be found in Ref. \cite{tan2019}.

The immediate result of this model for this study is the probability of $K^0-K^{0'}$ oscillations in vacuum \cite{tan2019},
\begin{equation}\label{eq_prob}
P_{K^0K^{0'}}(t) = \sin^2(2\theta) \sin^2(\frac{1}{2}\Delta_{K^0K^{0'}} t)
\end{equation}
where $\theta$ is the $K^0-K^{0'}$ mixing angle and $\sin^2(2\theta)$ denotes the mixing strength of about $10^{-4}$, $t$ is the propagation time, $\Delta_{K^0K^{0'}} = m_{K^0_2} - m_{K^0_1}$ is the small mass difference of the two mass eigenstates of about $10^{-6}$ eV \cite{tan2019}, and natural units ($\hbar=c=1$) are used for simplicity. Note that the equation is valid even for relativistic kaons and in this case $t$ is the proper time in the particle's rest frame. There are actually two weak eigenstates of $K^0$ in each sector, i.e., $K^0_S$ and $K^0_L$ with lifetimes of $9\times10^{-11}$ s and $5\times10^{-8}$ s, respectively. Their mass difference is about $3.5 \times 10^{-6}$ eV very similar to $\Delta_{K^0K^{0'}}$, which makes one wonder if the two mass differences and even the $CP$ violation may originate from the same source.

For kaons that travel in the thermal bath of the early universe, each collision or interaction with another particle will collapse the oscillating wave function into a mirror eigenstate. In other words, during mean free flight time $\tau_f$ the $K^0-K^{0'}$ transition probability is $P_{K^0K^{0'}}(\tau_f)$. The number of such collisions will be $1/\tau_f$ in a unit time. Therefore, the transition rate of $K^0-K^{0'}$ with interaction is \cite{tan2019},
\begin{equation}\label{eq_prob2}
\lambda_{K^0K^{0'}} = \frac{1}{\tau_f}\sin^2(2\theta) \sin^2(\frac{1}{2}\Delta_{K^0K^{0'}} \tau_f).
\end{equation}
Note that the Mikheyev-Smirnov-Wolfenstein (MSW) matter effect \cite{wolfenstein1978,mikheev1985}, i.e., coherent forward scattering that could affect the oscillations is negligible as the meson density is very low when kaons start to condensate from the QCD plasma (see more details for in-medium particle oscillations from Ref. \cite{tan2019a}), and in particular, the QCD phase transition is most likely a smooth crossover \cite{hotqcdcollaboration2014,fang2019}.

It is not very well understood how the QCD symmetry breaking or phase transition occur in the early universe, which will be discussed in detail in the next section. Let us suppose that the temperature of QCD phase transition $T_c$ is about 150 MeV and a different value (e.g., 200 MeV) here does not affect the following discussions and results. At this time only up, down, and strange quarks are free. It is natural to assume that strange quarks become confined first during the transition, i.e., forming kaon particles first instead of pions and nucleons. A better understanding of this process is shown in the next section. As a matter of fact, even if they all form at the same time, the equilibrium makes the ratio of nucleon number to kaon number 
\begin{equation}
\frac{n_N}{n_K} \simeq (\frac{m_N}{m_K})^{3/2} \exp (-(m_N-m_K)/T_c) \sim 0.1
\end{equation}
very small due to the fact that kaons are much lighter than nucleons.

Once neutral kaons are formed, they start to oscillate by participating in the weak interaction with cross section of $\sigma_{\text{EW}} \sim G^2_F T^2$ where $G_F = 1.17 \times 10^{-5} $ GeV$^{-2}$ is the Fermi coupling constant. Then one can estimate $K^0$'s thermally averaged reaction rate over the Bose-Einstein distribution,
\begin{eqnarray}
\Gamma &=& \frac{g}{(2\pi)^3}\int^{\infty}_0 d^3p f(p) \sigma_{\text{EW}} \frac{p}{m} \nonumber	\\
&=& \frac{g}{2\pi^2} \frac{G^2_F T^2}{m}\int^{\infty}_0 dp \frac{p^3}{\exp(\sqrt{p^2+m^2}/T)-1} \label{eq_fo}
\end{eqnarray}
where $g=2$ for both $K^0_S$ and $K^0_L$, $m$ is the mass of kaons, and $T$ is the temperature. The expansion rate of the universe at this time can be estimated to be $H \sim T^2_{\text{MeV}}$ s$^{-1}$ where $T_{\text{MeV}}$ is the temperature in unit of MeV. The condition for $K^0$ to decouple from the interaction or freeze out is $\Gamma / H < 1$. It can be easily calculated from Eq. (\ref{eq_fo}) that the freezeout occurs at $T_{fo}=100$ MeV. This means that kaon oscillations have to operate between $T_c=150$ MeV and $T_{fo}=100$ MeV. And fortunately the $K^0$ mesons have long enough lifetime (compared to the weak interaction rate) for such oscillations and BAU to occur during this temperature range.

For the standard constraint on the mirror-to-normal matter temperature ratio of $x=T'/T < 1/2$ \cite{kolb1985,hodges1993} that will be discussed further in Sec. \ref{origin}, the two oscillation steps of $K^{0'}\rightarrow K^{0}$ and $K^0\rightarrow K^{0'}$ will be decoupled in a similar way as the $n-n'$ oscillations discussed in Ref. \cite{tan2019}. Using a typical weak interaction rate $\lambda_{\text{EW}} = 1/\tau_f = G^2_F T^5 \sim T^5_{\text{MeV}}$ s$^{-1}$ and the age of the universe $t = 0.3/T^2_{\text{MeV}}$ s during this period of time, one can get the final-to-initial $K^{0}$ abundance ratio in the ordinary sector for the second step,
\begin{eqnarray} \label{eq_eps}
\frac{X^{f}}{X^{i}} &=& \exp(-\int P_{K^0K^{0'}}(\tau_f) \lambda_{\text{EW}} dt) \nonumber \\
&= & \exp(-5\times10^{28}\sin^2(2\theta)(\frac{\Delta_{K^0K^{0'}}}{\text{eV}})^2 \int^{T_{fo}}_{T_c} d(\frac{1}{T^7_{\text{MeV}}}) ) \nonumber \\
&=& 1-0.05 \equiv 1-\epsilon \label{eq_osc1}
\end{eqnarray}
and the first step is negligible due to the much faster expansion rate of the universe \cite{tan2019}.

However, $K^0_S$ has a lifetime of $9\times10^{-11}$ s that is comparable to the weak interaction rate at such temperatures. Owing to this, only one third of $K^0_S$ particles participate in the oscillations while the other two thirds decay to pions. In contrast to $K^0_S$, $K^0_L$ mesons have a much larger lifetime ($5\times10^{-8}$ s) and hence almost all of them take part in the oscillations. Considering the above correction, the final-to-initial $K^{0}$ abundance ratio in the normal world due to oscillations becomes,
\begin{equation}
\frac{X^{f}}{X^{i}} = 1 - \frac{2}{3}\epsilon.
\end{equation}
The $CP$ violation amplitude in SM is measured as $\delta=2.228\times10^{-3}$ \cite{particledatagroup2018} so that $\delta^2 \sim 5\times10^{-6}$ and the oscillation probability ratio can be estimated as $P_{K^0K^{0'}} / P_{\bar{K}^0\bar{K}^{0'}} \sim 1-\delta^2$. Then the net $K^0$ fraction can be obtained as follows,
\begin{equation} \label{dkk}
\frac{\Delta X_{K^0\bar{K}^0}}{X_{K^0\bar{K}^0}} \equiv \frac{X_{K^0} - X_{\bar{K}^0}}{X_{K^0} + X_{\bar{K}^0}} = \frac{1}{3}\epsilon \delta^2 \sim \frac{25}{3}\times 10^{-8}
\end{equation}

If the excess of $K^0 (d\bar{s})$ generated above can survive by some B-violation process, i.e., dumping $\bar{s}$ quarks and leaving $d$ quarks to form nucleons in the end, then assuming that half of strange quarks condensate into $K^0_{L,S}$ (with the other half in $K^{\pm}$) we will end up with a net baryon density of $n_B/s = 5.6\times 10^{-10}$ that essentially gives the sum of the observed baryon and dark matter.
In the next section, we will demonstrate how such a B-violation process could occur during the QCD phase transition.

In Eq. (\ref{eq_eps}) the mixing strength $\sin^2(2\theta) \sim 10^{-4}$ and the mass splitting parameter $\Delta_{K^0K^{0'}} \sim 10^{-6}$ eV are estimated from $n-n'$ oscillations in Ref. \cite{tan2019} assuming that the single-quark mixing strength is similar and the mass splitting parameter is scaled to the particle's mass. Unfortunately, these estimates are still fairly rough as the neutron lifetime measurements have not yet constrained the oscillation parameters well \cite{tan2019} resulting in a factor of $\sim 10$ uncertainty in $\epsilon$ of Eq. (\ref{eq_eps}). On the other hand, the observed baryon asymmetry can be used to constrain these parameters under the new mechanism, i.e., $\epsilon = 0.05$ or $\sin^2(2\theta)\Delta_{K^0K^{0'}}^2 = 10^{-16}$ eV$^2$. Remarkably, such parameters are consistent with the neutron lifetime experiments and the origin of dark matter under the new model (see more discussions in Sec. \ref{origin}). More detailed studies of the mirror mixing parameters under the context of the CKM matrix and proposed laboratory measurements can be found in a separate paper \cite{tan2019d}.

\section{QCD symmtry breaking transition and other oscillations\label{QCD}}

A massless fermion particle's chirality or helicity has to be preserved, i.e., its left- and right-handed states do not mix \cite{weinberg1995}. This is essentially also true for extremely relativistic massive particles as required by special relativity. Therefore the global flavor chiral symmetry of $SU(2)_L\otimes SU(2)_R$ for the family of up and down quarks is very good as their masses are so tiny compared to the QCD confinement energy scale.

Under strong interactions like QCD, the non-vanishing vacuum expectation value of quark condensates can lead to spontaneous symmetry breaking (SSB) by mixing left- and right-handed quarks in the mass terms. The resulting pseudo-Nambu-Goldstone bosons (pNGB) and Higgs-like field will manifest as light bound states of quark condensates. For example, the approximate $SU(2)_L\otimes SU(2)_R$ chiral symmetry is spontaneously broken into $SU(2)_V$, i.e., the isospin symmetry at low energies in QCD, which can be described under an effective theory of the so-called $\sigma$-model \cite{weinberg1995}. In this case, the lightest isoscalar scalar $\sigma$ or $f_0(500)$ meson with mass of $\sim 450$ MeV serves as the quark condensate for SSB \cite{pelaez2016}, a similar role to Higgs in electroweak SSB. The resulting pNGB particles are the three lightest pseudoscalar mesons ($\pi^{\pm}$ and $\pi^{0}$). The Lagrangian for the matter part with omission of gauge fields and Higgs-like parts can be written as,
\begin{equation}
L_{matter} = \bar{q}^a_L(i\gamma^{\mu}D_{\mu})q^a_L + \bar{q}^a_R(i\gamma^{\mu}D_{\mu})q^a_R - m_a(\bar{q}^a_L q^a_R + \bar{q}^a_R q^a_L)
\end{equation}
where the left- and right-handed quark fields $q_{L/R}$ are summed over the flavor index $a$. The non-vanishing mass terms can mix left- and right-handed states and hence explicitly break the chiral symmetry.

There is actually an extra global symmetry of $U(1)_L\otimes U(1)_R$ in the above QCD system before the SSB, where the $U(1)_{L+R}$ symmetry is conserved and manifests as baryon conservation in QCD while the axial part $U(1)_{L-R}$ or $U(1)_A$ is explicitly broken by the axial current anomaly, resulting in a $CP$ violating term in the Lagrangian involving gauge field $G$,
\begin{equation}
L_{\theta} = \frac{\theta g^2_s}{32\pi^2}G\cdot\tilde{G}
\end{equation}
with $\theta$ modified by the Yukawa mass matrices for quarks as the physical strong $CP$ phase $\bar{\theta}=\theta - \text{arg det}(\prod_a m_a)$. This leads to the long-standing so-called $U(1)_A$ and strong $CP$ puzzles in particle physics \cite{bigi2009} as the $\bar{\theta}$ parameter has to be fine-tuned to zero or at least $\leq 10^{-9}$ to be consistent with experimental constraints of the neutron electric dipole moment \cite{pendlebury2015}.

In the scheme of $1/N$ expanded QCD, Witten using a heuristic method \cite{witten1979} discovered an interesting connection to the $\eta'$ meson as a possible pNGB to solve the $U(1)_A$ or strong $CP$ problem although the $\eta'$ mass (958 MeV) seems to be too high for the above chiral SSB. The good Witten-Veneziano relation for obtaining the $\eta'$ mass under such an approach \cite{witten1979,veneziano1979} indicates some validity of the idea. In addition, it gives the correct QCD transition scale of about 180 MeV and relates the $\eta'$ mass to the interesting topological properties of QCD \cite{witten1979,veneziano1979}.

At a little earlier time, Peccie and Quinn \cite{peccei1977,peccei1977a} conjectured a so-called $U(1)_{PQ}$ axial symmetry to solve the $U(1)_A$ problem by dynamically canceling the axial anomaly with an imagined ``axion'' field. Here we could combine the two brilliant ideas and find the clue for solving the problem as shown below.

The key is to realize that the QCD symmetry breaking transition can be staged as shown in Table \ref{tab_qcd}. That is, we could have a strange quark condensation first leading to an SSB at a higher energy scale and then the normal $SU(2)$ chiral SSB at slightly lower energy. At the early stage, it is the strange $U(1)$ (i.e., $U_s(1)$) symmetry that gets spontaneously broken. The $U_s(1)_{L+R}$ is kept as strange number conservation in QCD that will then be broken by the electroweak force while the other global $U_s(1)_{L-R}$ symmetry is broken by mixing left- and right-handed strange quarks in the mass term. At the same time the $SU(3)$ flavor symmetry of (u,d,s) quarks is broken into $SU(2)$ of (u,d) quarks with five pNGB particles of $K^{\pm}$, $K^{0}_{L,S}$, and $\eta$ (more exactly $\eta_8$). The broken $U_s(1)_{L-R}$ or $U_s(1)_A$ gives another pNGB, i.e., $\eta'$ (more exactly $\eta_1$ with quark configuration of $u\bar{u}+d\bar{d}+s\bar{s}$), as Witten suspected. The Higgs-like particle leading to this SSB is the scalar singlet $f_0(980)$ meson with mass of $990$ MeV \cite{particledatagroup2018} that is perfectly compatible with the seemingly heavy $\eta'$.

\begin{table}
\caption{\label{tab_qcd} Possible stages of QCD spontaneous symmetry breaking or phase transitions are shown. Candidates of Higgs-like and pNGB particles are taken from the compilation of Particle Data Group \cite{particledatagroup2018}. The major oscillations of neutral condensates and non-perturbative processes at each stage are listed as well.}
\begin{ruledtabular}
\begin{tabular}{c c L c c L}
SSB stages & $(u,d)$ & $s\bar{s}$ & $c\bar{c}$ & $b\bar{b}$ & $t\bar{t}$ \\
\hline
Higgs-like & $\sigma/f_0(500)$ & $f_0(980)$ & $\chi_{c0}(1P)$ & $\chi_{b0}(1P)$ & Higgs \\
\hline
Broken Symm. & chiral $SU(2)$ & $U_s(1)_A$ and $SU(3)\rightarrow SU(2)$ & $U_c(1)_A$ & $U_b(1)_A$ & $U_t(1)_A$ and EW \\
\hline
pNGB & $\pi^{\pm},\pi^0$ & $\eta_1(\eta')$ and $K^{\pm},K^0_{L,S},\eta_8(\eta)$ & $\eta_c(1S)$  & $\eta_b(1S)$  & $\eta_t(1S)$? \\
\hline
Oscillations & $n-n'$ & $K^0-K^{0'}$ & $D^0-D^{0'}$ & $B^0-B^{0'}$ & $H-H'$ \\
\hline
Non-perturbative & & s-quarkiton & c-quarkiton & b-quarkiton & t-quarkiton and sphaleron
\end{tabular}
\end{ruledtabular}
\end{table}

The $U_s(1)_{A}$ symmetry has all the desired necessary features of the arbitrary $U(1)_{PQ}$ axial symmetry conjectured by Peccie and Quinn \cite{peccei1977,peccei1977a}. That is, SSB of $U_s(1)_{A}$ due to strange quark condensation provides a Higgs-like field ($f_0(980)$) and a pNGB ($\eta_1$) that can dynamically drive the $U(1)_A$ axial anomaly and the $\bar{\theta}$ parameter to zero and therefore solving the strong $CP$ problem. The imagined ``axion'' from SSB of the Peccie-Quinn symmetry \cite{peccei2008} is not needed and the problem can be solved within the framework of SM without new particles.

However, such a solution does not seem to provide a B-violation mechanism for solving the BAU problem as $U_s(1)_{L+R}$ or strange number is conserved. Another key insight related to the non-perturbative effects and topological structures of QCD and SM will be discussed below.

The work of 't Hooft \cite{thooft1976,thooft1976a} interpreted the $U(1)_{A}$ anomaly in the chiral SSB as the topological effects in QCD and introduced the so-called $\theta$-vacua between which tunneling occurs via instantons non-perturbatively although such quantum tunneling effects are extremely suppressed. It is actually this kind of non-trivial $\theta$-vacuum structure and instanton-like gauge field solutions leading to the desired $B$-violation in SM. Below we provide a brief review of the known electroweak sphaleron under gauge SSB and then propose a new type sphaleron-like process using the above-discussed dynamic SSB on global symmetries.

A saddle-point gauge field solution called ``sphaleron'' in the electroweak interaction of $SU(2)_L \otimes U(1)_Y$ was first discovered in 1984 by Klinkhamer and Manton \cite{klinkhamer1984} that inspired various electroweak baryogenesis models later. Finite temperature effects considered by Ref. \cite{kuzmin1985} make the sphaleron-like process rate high enough for $B$-violation around or above the electroweak phase transition energy scale. Recently an $SU(3)$ sphaleron has been proposed and calculated \cite{klinkhamer2005,klinkhamer2017} and could be related to the non-Abelian chiral anomaly \cite{bardeen1969}.

The nontrivial vacuum structure in gauge theories can be characterized by the Chern-Simons integer or the winding number $N_{CS}$
and transitions between topologically inequivalent vacuum configurations can then be denoted by the integer Pontryagin index or the topological charge,
\begin{equation}
Q \equiv \Delta N_{CS} = \frac{g^2}{32\pi^2} \int d^4x G\cdot\tilde{G}.
\end{equation}

The electroweak sphaleron is associated with the SSB of the electroweak gauge symmetry $SU(2)$ and the global $B$ and $L$ anomalies of $SU(2)^2 U(1)_B$ and $SU(2)^2 U(1)_L$, respectively. The corresponding anomalous baryon and lepton number currents can be written as,
\begin{equation}
\partial^{\mu} J^{B}_{\mu} = \partial^{\mu} J^{L}_{\mu} = \frac{g^2 N_g}{16\pi^2} G\cdot\tilde{G}
\end{equation}
and therefore the baryon and lepton number conservation is violated for a topological transition as follows,
\begin{equation}
\Delta B = \Delta L = 2N_g \Delta N_{CS}
\end{equation}
where $N_g=3$ is the number of generations. The sphaleron sits at the top of the energy barrier between two adjacent vacuum configurations with $\Delta N_{CS} = 1/2$ and hence it involves nine quarks and three leptons (all left-handed) from each generation and violates $B$ and $L$ numbers by three units (i.e., $N_g$) while conserving $B - L$ at the same time. The sphaleron energy can be estimated as \cite{klinkhamer1984},
\begin{equation}
E_{s} \sim M_W / \alpha \sim 10 \text{ TeV}
\end{equation}
which essentially defines the height of the barrier between topologically disconnected vacua.

Now the question becomes if there is a similar sphaleron-like process that could occur at the energy scale of the QCD phase transition. The answer is very likely. There could be a similar saddle-point solution when the QCD gauge fields are included with a dynamic SSB on global symmetries, we will call it ``quarkiton'' to distinguish from sphaleron for the electroweak gauge SSB.

The quarkiton process is assumed to be associated with the strange quark condensation and the strange chiral $U_s(1)_A$ SSB as discussed above. As such, it is related to the strange chiral anomaly of $SU(3)_c^2 U_s(1)_A$ under QCD with an anomalous chiral current for strange quarks expressed by its divergence,
\begin{equation}
\partial^{\mu} J^{5s}_{\mu} = \frac{g^2 N_c}{16\pi^2} G\cdot\tilde{G}
\end{equation}
and the strange chirality violation can be obtained as follows,
\begin{equation}
\Delta S_c = 2N_c \Delta N_{CS}
\end{equation}
where $N_c=3$ is the quark color degree of freedom. This chirality violation requires three strange quarks of the same chirality to form the quarkiton at the top of the energy barrier between two neighboring QCD vacuum configurations with $\Delta N_{CS} = 1/2$ like the sphaleron.
When the electroweak gauge symmetry is also considered, the chiral anomaly of $SU(2)^2_L U(1)_A$ within the 2nd generation of quarks and leptons provide additional selection rules of $\Delta B = \Delta L = 2 \Delta N_{CS} = 1$ for the quarkiton. For the full SM gauge theory, therefore, it is natural to construct the quarkiton as a B and L violating process (by one unit for each) involving three strange quarks and three leptons in the same generation like the following,
\begin{equation}\label{eq_sss}
sss+\mu^+ \nu_{\mu}\nu_{\mu} \Leftrightarrow \text{Quarkiton} \Leftrightarrow \bar{s}\bar{s}\bar{s}+\mu^- \bar{\nu}_{\mu}\bar{\nu}_{\mu}
\end{equation}
where all of quarks and leptons are left-handed, three strange quarks ensure a color singlet, and the overall $B-L$ is conserved. In particular, a quarkiton is configured to be a neutral singlet under the SM gauge symmetry of $SU(3)_c \otimes SU(2)_L \otimes U(1)_Y$. A complete topological transition of $\Delta N_{CS} = 1$ via quarkiton can be described by Eq. (\ref{eq_sss}) as follows: the lhs particles excited out of one vacuum configuration form the quarkiton over the barrier and then decay to the corresponding anti-particles on the rhs of Eq. (\ref{eq_sss}) with respect to the next vacuum configuration.

To estimate the quarkiton energy, we apply SSB on the global flavor symmetry $SU(3)$ of (u,d,s) quarks and the chiral strange $U_s(1)$ instead of the gauge symmetries as used in sphaleron calculations \cite{klinkhamer1984,klinkhamer2017}. We can derive a similar saddle-point solution with its energy related to the pNGB particles of quark condensates instead of the elementary gauge bosons. In particular, the quarkiton energy can be related to the kaon mass and the kaon-quark coupling as follows,
\begin{equation}
E_{q} \sim m_K / \alpha_{Kqq} \sim 0.5 \text{ GeV}
\end{equation}
where the kaon-quark coupling $\alpha_{Kqq} = g^2_{Kqq}/4\pi \sim 1$ is inferred from the observed pion-nucleon coupling constant $g_{\pi NN} = 13.4$, the corresponding pion-quark coupling constant $g_{\pi qq} \approx g_{\pi NN}/(3g_A) \approx 3.6$, and $\alpha_{\pi qq} = g^2_{\pi qq}/4\pi \sim 1$. So the quarkiton energy is on the same order of the kaon mass $m_K \sim 0.5$ GeV and close to the energy scale of 0.2 GeV for the strange quark condensation or phase transition. Such a low energy barrier ensures that the quarkiton transition rate is high enough for B-violation around the QCD phase transition energy scale.

Such a quarkiton process can help solve the BAU problem under the scenario of $K^0-K^{0'}$ oscillations discussed in the previous section. Like the electroweak transition \cite{kajantie1996}, the QCD phase transition in the early universe is most likely a smooth crossover \cite{hotqcdcollaboration2014,fang2019} and the extra $K^{0}$ ($d\bar{s}$) particles will be deconfined back into free down and anti-strange quarks. That is, the extra down quarks from $K^{0}$ can be saved once all the extra anti-strange quarks are converted to strange quarks via the quarkiton process and then condensate again into mesons. Half of the saved down quarks are subsequently transitioned to up quarks by the electroweak interaction. When the next stage QCD phase transition (i.e., the chiral $SU(2)$ SSB) occurs at possibly around $T=100-150$ MeV or temperatures mostly overlapped with the s-quark condensation process for a smooth phase transition crossover, these extra up and down quarks will condensate into protons and neutrons forming the initial baryon content of the universe. The net effect after all these processes for one $K^{0}$ ($d\bar{s}$) excess is,
\begin{equation}\label{eq_mu}
d + \bar{s} \rightarrow \frac{1}{6}p + \frac{1}{6}n + \frac{1}{6} e^- + \frac{1}{6}\bar{\nu}_e + \frac{1}{3}\nu_{\mu}.
\end{equation}

During the strange quark condensation, kaons are the lightest strange mesons. So it is safe to assume that about half of strange quarks condensate into $K^0$ while the other half into $K^{\pm}$. Before condensation the (anti-)strange quark number to entropy density ratio is $n_{s\bar{s}}/s=4\times 10^{-2}$ owing to an effective number of relativistic degrees of freedom $g_* = 61.75$ during this stage. Taking into account the oscillation result from Eq. (\ref{dkk}) one can obtain a net baryon-number-to-entropy density ratio of $n_B/s = 5.6\times 10^{-10}$. Considering that most of the baryon excess generated above will be converted to mirror baryons subsequently via $n-n'$ oscillations \cite{tan2019} and today's observed dark/mirror-to-baryon ratio is 5.4, the eventual leftover baryons in the normal sector will be $n_B/s = 8.7\times 10^{-11}$ that agrees very well with the observed value.

Note that $B-L$ is conserved at the end of net baryon generation from (\ref{eq_mu}) with extra amount of $\nu_{\mu}$ equal to the net baryon number. The fate of these and other neutrinos and their effects on the thermal evolution of the universe will be discussed in the next section.

Now one may wonder if a similar quarkiton process and SSB could also operate earlier at higher temperatures for charm, bottom, and even top quark condensation. Interestingly, analogous to the strange quarkiton process, the following could be conceived to occur at different condensation stages for c-, b-, and t- quarks, respectively,
\begin{eqnarray}
ccc +\mu^- \mu^-\bar{\nu}_{\mu} \Leftrightarrow &\,\text{Quarkiton}\,& \Leftrightarrow \bar{c}\bar{c}\bar{c}+\mu^+ \mu^+\nu_{\mu} \\
bbb +\tau^+ \nu_{\tau}\nu_{\tau} \Leftrightarrow &\,\text{Quarkiton}\,& \Leftrightarrow \bar{b}\bar{b}\bar{b}+\tau^- \bar{\nu}_{\tau}\bar{\nu}_{\tau} \\
ttt +\tau^- \tau^-\bar{\nu}_{\tau} \Leftrightarrow &\,\text{Quarkiton}\,& \Leftrightarrow \bar{t}\bar{t}\bar{t}+\tau^+ \tau^+\nu_{\tau}
\end{eqnarray}
where the SM gauge singlet configuration is required for all quarkitons. The Higgs-like candidates could be $\chi_{c0}(1P)$ for c-quark condensation and 
$\chi_{b0}(1P)$ for b-quark condensation with the possible pNGB particles of $\eta_c(1S)$ and $\eta_b(1S)$ for breaking the corresponding $U_c(1)_A$ and $U_b(1)_A$ symmetries, respectively, as shown in Table \ref{tab_qcd}.

Another interesting idea could be conceived from the coincident energy scale of t-quark condensation and electroweak phase transition. That is, the actual Higgs could be a bound state of top quark condensate that breaks both the global QCD top flavor $U_t(1)_A$ and the electroweak gauge symmetries at the same time by giving mass to all the fermion particles and defining the SM vacuum structure. The subsequent b-, c-, s- quark condensation and SSB transitions just modify the QCD vacuum structure further. Together with evidence of similar $K^0$ mass differences due to $CP$ violation and mirror splitting as discussed earlier, one may wonder if at the scale of $T_{\text{EW}}$ the top quark condensation could also break the degeneracy of normal and mirror worlds and cause the $CP$ violation at the same time.

These phase transition processes can lead to more particle oscillations between the normal and mirror sectors from $D^0$, $B^0$, and Higgs during the c-, b-, and t-quark condensation phases, respectively. For Higgs with $\Delta_{HH'} \sim 10^{-4}$ eV and $\sin^2(2\theta) \sim 10^{-4}$ (as a t-quark condensate) \cite{tan2019}, one can get a small oscillation parameter of $\epsilon(HH') \sim 10^{-18}$ at $T_c = 100$ GeV from Eq. (\ref{eq_osc1}). For $D^0$ with $\Delta_{D^0D^{0'}} \sim 10^{-6}$ eV and $\sin^2(2\theta) \sim 10^{-4}$ \cite{tan2019}, we can estimate $\epsilon(D^0D^{0'}) \sim 10^{-8}$ at $T_c = 1$ GeV from Eq. (\ref{eq_osc1}). Similarly, $\epsilon(B^0B^{0'}) \sim 10^{-13}$ at $T_c = 10$ GeV for $B^0$ with $\Delta_{B^0B^{0'}} \sim 10^{-5}$ eV and $\sin^2(2\theta) \sim 10^{-4}$ \cite{tan2019}. These oscillations are much weaker compared to the $K^0-K^{0'}$ oscillations and therefore they are negligible for the generation of BAU.

\section{Consistent origin of BAU and dark matter\label{origin}}

As demonstrated in the previous sections, $K^0-K^{0'}$ oscillations provide an intriguing mechanism for the generation of the matter-antimatter asymmetry. In combination with the $n-n'$ oscillations under the new model \cite{tan2019}, a consistent picture for the origin of both BAU and dark matter will be presented in this section.

Besides the two built-in model parameters of the mixing strength $\sin^2(2\theta)$ and the mass difference $\Delta$, a third cosmological parameter $x=T'/T$ has to be constrained for such oscillations to work. To be consistent with the results of the standard big bang nucleosynthesis (BBN) model, in particular, the well known primordial helium abundance, a strict requirement of $T'/T < 1/2$ at BBN temperatures \cite{kolb1985,hodges1993,foot2004,berezhiani2004} has to be met to ensure a slow enough expansion of the universe. Such a temperature condition can naturally occur after the early inflation and subsequent reheating \cite{kolb1985,hodges1993}. A typical ratio of $T'/T \sim 0.3$ is evidently supported in the studies of ultrahigh energy cosmic rays under the new mirror matter model \cite{tan2019b}. A better constraint on this parameter can probably be obtained from the thorough BBN simulations modified with the $n-n'$ oscillations of the new mirror matter model, which could potentially solve the primordial $^7$Li puzzle \cite{coc2013,coc2014a} as well.

As shown in the earlier discussions, the normal and mirror sectors do not exchange much via oscillations in the early universe, only on the order of $10^{-8}$ or less for $D^0$, $B^0$, and Higgs oscillations. The largest exchange of a few percents comes from $K^0-K^{0'}$ oscillations. Although the $n-n'$ oscillations \cite{tan2019} are more dramatic, the overall baryon density is too low at the moment and consequently the $n-n'$ induced exchange between the two sectors is much smaller. Therefore, the entropy of each sector is approximately conserved from the electroweak phase transition ($T = 100$ GeV) until after BBN. Meanwhile, the macroscopic asymmetry on the ratio of $T'/T$ is mostly preserved as well.

However, neutrino-mirror neutrino oscillations could become significant when the universe cools down to $T=0.8$ keV or about 20 days after the Big Bang assuming $\Delta^2_{\nu\nu'} \sim 10^{-18}$ eV$^2$ \cite{tan2019}. This can significantly change the entropy of each sector and also the temperatures of normal and mirror neutrinos. On the other hand, the normal and mirror gamma temperatures should stay intact since neutrinos have decoupled long before this moment. When we discuss the baryon-to-entropy ratio of $n_B/s$ the traditional entropy definition is used by ignoring the entropy changes due to possible $\nu-\nu'$ oscillations. Notwithstanding, a new understanding of the nature of neutrinos and mirror neutrinos in the extended Standard Model with Mirror Matter (SM$^3$) predicts that such $\nu-\nu'$ oscillations are not possible \cite{tan2019e}.

Only the criterion of $T'/T < 1/2$ is needed for the studies in this paper. To better illustrate the process, however, we use $T'/T = 1/3$ as an example with the sequence of the events listed in Table \ref{tab_osc}. Mirror oscillations for both $K^{0'}$ and $n'$ occur first when the normal sector is still above the QCD phase transition temperature making its effective number of relativistic degrees of freedom $g_*$ much larger. These early oscillation processes contribute little as their oscillation parameter of $\epsilon = \int P(\tau_f) \lambda dt$ is greatly suppressed by a factor of $(T'/T)^2\sqrt{g_*(T')/g_*(T)}$ \cite{tan2019}.

\begin{table}
\caption{\label{tab_osc} The sequence of events in the early universe is listed during the period of $K^0-K^{0'}$ and $n-n'$ oscillations using $T'/T = 1/3$ as an example. The QCD phase transition temperature is assumed to be 150 MeV.}
\begin{ruledtabular}
\begin{tabular}{c c p{12cm}}
$T$ [MeV] & $T'$ [MeV] & events \\
\hline
450 & 150 & start of mirror s-quark condensation and suppressed $K^{0'} \rightarrow K^{0}$ oscillations; start of mirror nucleon formation and suppressed $n' \rightarrow n$ oscillations (possibly slightly later) \\
300 & 100 & end of suppressed $K^{0'} \rightarrow K^{0}$ oscillations\\
210 & 70 & peak of suppressed $n' \rightarrow n$ oscillations \\
150 & 50 & start of normal s-quark condensation and $K^{0} \rightarrow K^{0'}$ oscillations; start of normal nucleon formation, $n \rightarrow n'$ oscillations, and generation of matter-antimatter asymmetry (possibly slightly later)\\
100 & 33 & end of $K^{0} \rightarrow K^{0'}$ oscillations \\
70 & 23 & peak of $n \rightarrow n'$ oscillations\\
60 & 20 & major $n \rightarrow n'$ conversion peak done \\
10 & 3 & tailing of $n \rightarrow n'$ oscillations; final dark(mirror)-to-baryon matter ratio  \\
1 & 0.3 & normal weak interaction decoupling \\
0.3 & 0.1 & start of mirror BBN \\
0.15 & 0.05 & mirror helium formed; mirror neutrons gained from $n \rightarrow n'$ oscillations \\
0.1 & 0.03 & start of normal BBN \\
0.05 & 0.017 & normal helium formed; low energy normal neutrons gained from $n' \rightarrow n$ oscillations resulting the destruction of $^7$Be \\
0.8 keV & 0.3 keV & start of $\nu-\nu'$ oscillations \cite{tan2019} or no $\nu-\nu'$ oscillations in SM$^3$ \cite{tan2019e} \\
\end{tabular}
\end{ruledtabular}
\end{table}

When the $K^{0} \rightarrow K^{0'}$ oscillations operate between $T=100-150$ MeV, initial matter-antimatter asymmetry is generated in the normal sector as discussed earlier while the excess of $\bar{K}^{0'}$ in the mirror sector will quickly decay into mirror pions at much lower mirror temperatures. Therefore, nearly all the initial baryon asymmetry originates from the normal sector and the mirror sector contributes little to the net baryon content in the beginning.

Possibly slightly after the inception of $K^{0} \rightarrow K^{0'}$ oscillations, the $n \rightarrow n'$ oscillations start to convert  the initial net baryons into mirror baryons. 
For a likely smooth crossover of QCD phase transition \cite{hotqcdcollaboration2014,fang2019}, the two oscillation processes probably overlap over a large temperature range (e.g., between 100 and 150 MeV). The peak of $n \rightarrow n'$ oscillations occurs at about 70 MeV well after the end of $K^{0} \rightarrow K^{0'}$ oscillations. The $n-n'$ oscillation rate drops quickly below $T=60$ MeV whereas $n-n'$ oscillations still keep a very small exchange rate between the two sectors even at temperatures below 10 MeV. The final mirror-to-normal baryon ratio eventually becomes about 5.4 as the observed dark-to-baryon ratio.

Once the mirror BBN starts, most of mirror neutrons will be fused into mirror helium. Instead of having mirror neutrons depleted as in standard BBN calculations, $n-n'$ oscillations will keep an appreciable $n'$ abundance in the mirror sector. The normal BBN then follows and most of normal neutrons are fused into normal helium. At this moment, the reverse $n' \rightarrow n$ oscillations will feed the normal sector with more neutrons at lower energies. These additional low energy neutrons will help destroy the extra $^7$Be formed earlier and potentially solve the primordial $^7$Li problem \cite{coc2013,coc2014a}. In particular, lower energy neutrons can make the $^7$Be destruction rate much higher than the n+p fusion rate \cite{kusakabe2014} and therefore alleviate the issue of lithium-deuterium anti-correlation \cite{coc2015}.

\begin{figure}
\includegraphics[scale=0.8]{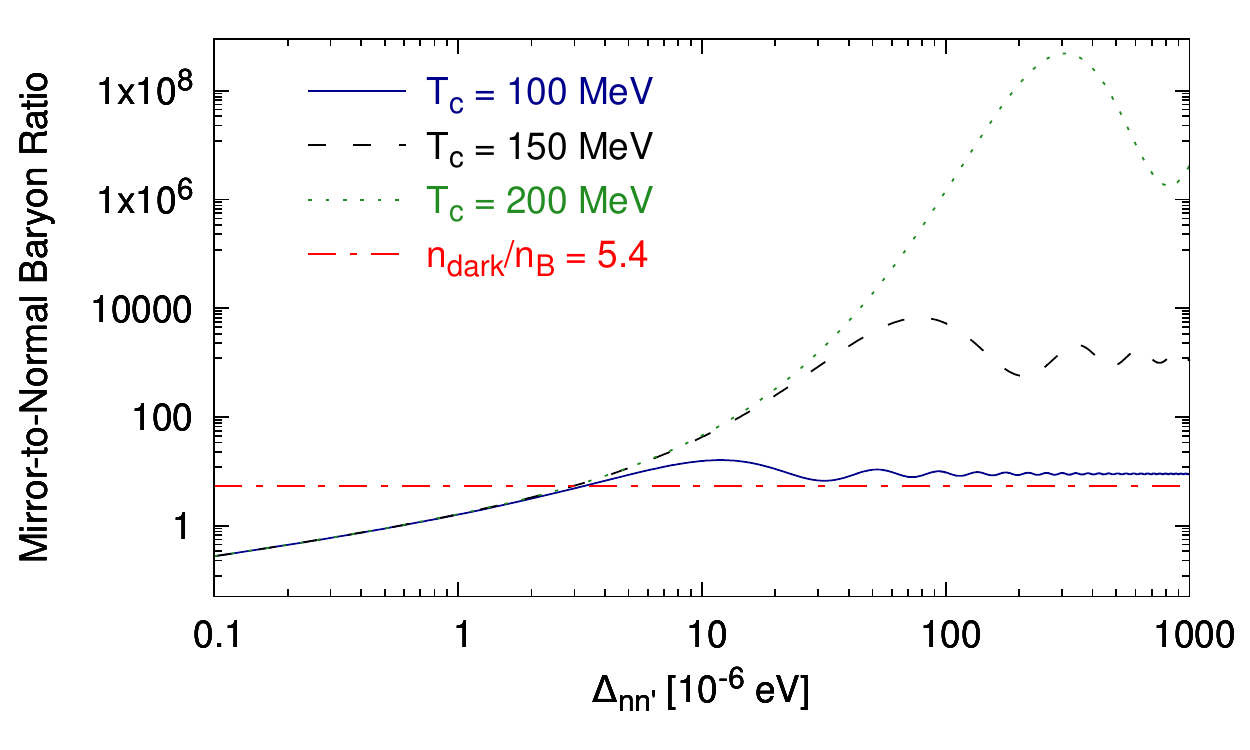}
\caption{\label{fig_ratio} The mirror-to-normal baryon ratio is shown as function of the mass difference $\Delta_{nn'}$ assuming the mixing strength $\sin^2(2\theta) = 2\times 10^{-5}$. The QCD phase transition temperature is varied to be $T_c=100, 150, 200$ MeV, respectively.}
\end{figure}

\begin{figure}
\includegraphics[scale=0.8]{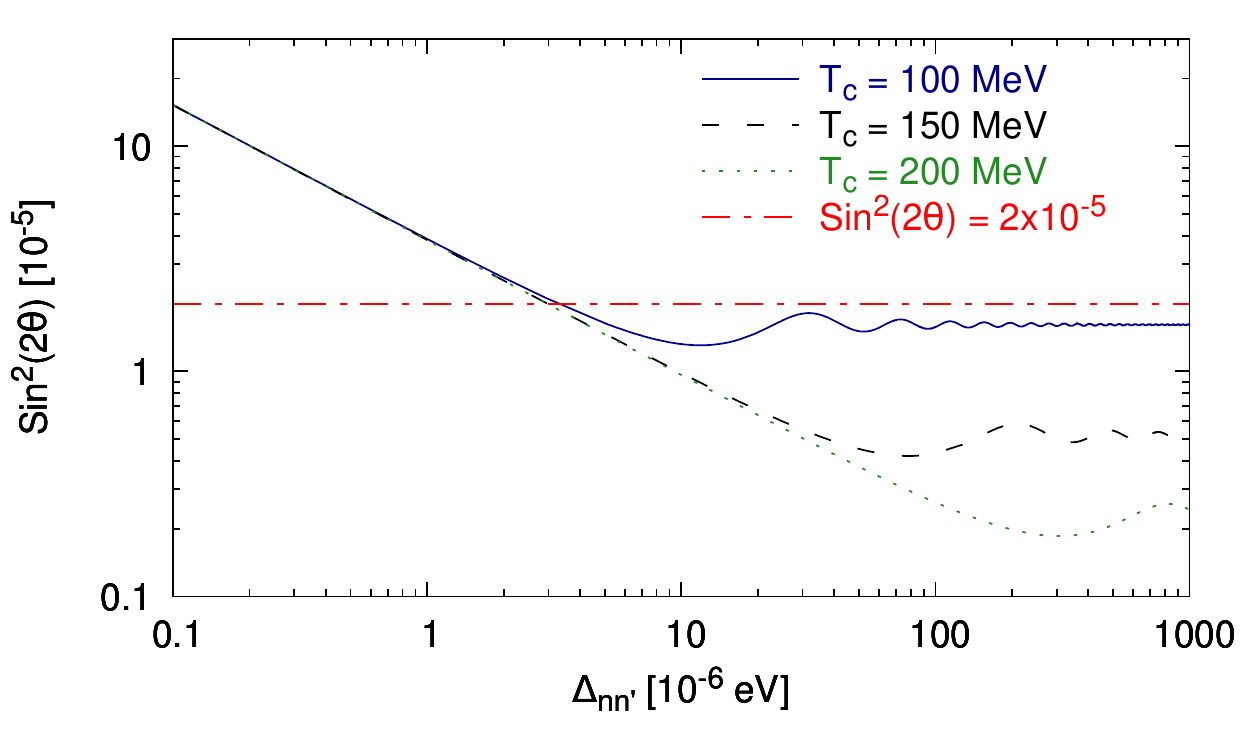}
\caption{\label{fig_mix} The mixing strength $\sin^2(2\theta)$ is shown as function of the mass difference $\Delta_{nn'}$ assuming a mirror-to-normal baryon ratio of 5.4. The QCD phase transition temperature is varied to be $T_c=100, 150, 200$ MeV, respectively.}
\end{figure}

Under the above consistent picture of particle oscillations, one can further examine the relations between the mixing strength $\sin^2(2\theta)$, the mass difference $\Delta_{nn'}$, the mirror-to-normal baryon ratio, and the QCD phase transition temperature $T_c$ using the framework developed in the original work of the new mirror matter model \cite{tan2019}. Figure \ref{fig_ratio} shows the mirror-to-normal baryon ratio as function of the mass difference for varied QCD phase transition temperatures and $\sin^2(2\theta)=2\times 10^{-5}$. Figure \ref{fig_mix} depicts the mixing strength vs the mass difference for three different QCD phase transition temperatures assuming a mirror-to-normal baryon ratio of 5.4.

For the most likely QCD phase transition temperature range (150-200 MeV) \cite{hotqcdcollaboration2014,fang2019} and the well observed dark-to-baryon ratio of 5.4, the $n-n'$ mass difference, as shown in Figs. \ref{fig_ratio}-\ref{fig_mix}, can be constrained as $\Delta_{nn'} = 10^{-6} - 10^{-5}$ eV by the uncertainty of the mixing strength $8\times 10^{-6} \leq \sin^2(2\theta) \leq 4\times 10^{-5}$ inferred from neutron lifetime measurements \cite{tan2019}. The best value of $\Delta_{nn'} = 3\times 10^{-6}$ eV corresponds to the mixing strength of $\sin^2(2\theta) = 2\times 10^{-5}$. Note that no upper limit on $\Delta_{nn'}$ can be set if the QCD phase transition temperature is somehow much lower (e.g., around 100 MeV). A detailed study of applying the new model to BBN may help further constrain these parameters. More neutron lifetime measurements with different magnetic traps can certainly provide a much more accurate value of $\sin^2(2\theta)$ and consequently better $\Delta_{nn'}$. And furthermore, it may also provide a way to pin down the QCD phase transition temperature.

\section{Conclusion\label{conclusion}}

Under the new mirror-matter model \cite{tan2019} and new understanding of possibly staged QCD symmetry breaking phase transitions, the long-standing BAU puzzle can be naturally explained with $K^0-K^{0'}$ oscillations that occur at the stage of strange quark condensation. A consistent picture of particle-mirror particle oscillations throughout the early universe is presented including a self-consistent origin of both BAU and dark matter. Meanwhile, the $U(1)_A$ or strong $CP$ problem in studies of particle physics is understood under the same framework. The connection between the $CP$ violation in SM and the normal-mirror mass splitting seemingly points to the same mechanism in new physics that needs to be explored in the future.
Non-perturbative processes via quarkitons at different quark condensation stages are proposed for B-violation and could be verified and further understood with calculations using the lattice QCD technique. More accurate studies on $K^0_{L,S}$ at the kaon production facilities, in particular, on the branching fractions of their invisible decays \cite{tan2019} that surprisingly are not constrained experimentally \cite{gninenko2015}, will better quantify the generation of baryon matter in the early universe. Future experiments at the Large Hadron Collider may provide more clues for such topological quarkiton processes and reveal more secrets in the SM gauge structure, the Higgs mechanism, and the amazing oscillations between the normal and mirror worlds.

\begin{acknowledgments}
This work is supported in part by the National Science Foundation under
Grant No. PHY-1713857 and the Joint Institute for Nuclear Astrophysics (JINA-CEE, www.jinaweb.org), NSF-PFC under Grant No. PHY-1430152.
\end{acknowledgments}

\bibliography{bau}

\end{document}